\documentclass[aps,prl,twocolumn,preprintnumbers,secnumarabic,amsmath,amssymb,nofootinbib,floatfix]{revtex4}
\usepackage{varioref,exscale,latexsym,amsmath,amssymb}
\usepackage{graphicx}
\usepackage{graphicx}% Include figure files
\usepackage{dcolumn}% Align table columns on decimal point
\usepackage{bm}% bold math

\def\beq{\begin{equation}}

\def\eeq{\end{equation}}

\def\beqa{\begin{eqnarray}}

\def\eeqa{\end{eqnarray}}

\begin{document}

\title{{\bf $\theta$ dependence of the deconfining phase transition in pure $SU(N_c)$ Yang-Mills theories }}

\medskip\
\author{Mohamed M. Anber}
\email[Email: ]{manber@physics.utoronto.ca}
\affiliation{Department of Physics,
University of Toronto\\
Toronto, ON, M5S1A7, Canada\\
}

\begin{abstract}
Recently, it has been conjectured that deconfining phase transition in $SU(N_c)$ pure Yang-Mills theories is continuously connected to a quantum phase transition in softly broken ${\cal N}=1$ super Yang-Mills on $R^{1,2} \times S^1$. We exploit this conjecture to study the strength of the transition  and  deconfining temperature as a function of the vacuum angle $\theta$ in pure Yang-Mills. We find that the transition temperature is a decreasing function of $\theta \in [0, \pi)$, in an excellent agreement with recent lattice simulations. We also predict that the transition becomes stronger for the same range of $\theta$, and comment on the $\theta$ dependence in the large $N_c$ limit. More lattice studies are required to test our predictions. 

\end{abstract}
%\vspace{0.2 in}
%\end{titlepage}
%\setcounter{page}{0}
%\newpage
\maketitle
%\documentstyle[12pt,epsfig]{article}
%\documentstyle[12pt,epsf,epsfig]{article}

%%%%%%%%%%%%%%%%%%%%%%%%%%%%%%%%%%%%%%%%

%%%%%%%%%%%%%%%%%%%%%%%%%%%%%%%%%%%%%%%%%%%%%%%%
%\section{Introduction}
%%%%%%%%%%%%%%%%%%%%%%%%%%%%%%%%%%%%%%%%%%%%%%%%

The presence of degenerate topological sectors in Yang-Mills theories entails  adding a term ${\cal L}_\theta=-i\frac{\theta}{32\pi^2g^2}F_{\mu\nu}^a\tilde F^{a\mu\nu}$ to the Lagrangian, where $\theta$ is the vacuum angle. In pure Yang-Mills or in the presence of massive fermions, both ${\cal C}$ and ${\cal CP}$ symmetries are lost. In $QCD$, the absence of ${\cal CP}$ violation puts stringent bound on the value of $\theta<10^{-10}$. However, at finite temperature and close to the deconfinement transition, $\theta$ can have spacetime dependence which gives rise to interesting physical phenomena in heavy ion collisions \cite{Kharzeev:2004ey}. On the other hand, studying the nature of the deconfinement transition in Yang-Mills theories is plagued by strongly coupled physics. All this makes it very necessary to move to a simpler theory that is under complete analytic control, yet resembles the original one. 

Recently, it has been argued that thermal phase transitions in pure Yang-Mills $SU(N_c)$ theories are continuously connected to quantum phase transitions in  Yang-Mills theories formulated on $R^{1,2}\times S^1$, with one adjoint Weyl fermion of mass $m$ and periodic boundary conditions   \cite{Poppitz:2012sw,Poppitz:2012nz}. The mass term is crucial since  the theory reduces to thermal pure gauge theory in the $m \rightarrow \infty$ limit, and to circle-compactified nonthermal ${\cal N}=1$ supersymmetric theory as $m \rightarrow 0$. At small compactification radius $L$, the theory is weakly coupled and  analytically tractable. Near $m=0$, the critical comactification radius $L^{\mbox{\scriptsize cr}}$ can be computed using reliable semi-classical methods. On the other hand, taking the limit $m \rightarrow \infty$, one can interpret $L^{\mbox{\scriptsize cr}}$ as the inverse critical temperature $T^{\mbox{\scriptsize cr}}$ of pure Yang-Mills.

At small $L$, the supersymmetric theory on $R^{1,2}\times S^1$, with a gaugino mass term, dynamically breaks down (due to adjoint Higgssing in the $S^1$ direction) to $U(1)^{N_c-1}$ at distances larger than $1/m_W$, where $m_W=2\pi/(L N_c)$ is the W-boson mass. Then, the  effective degrees of freedom  are $N_c-1$ dual photons $\vec \sigma$ (a $N_c-1$ dimensional vector) and $\vec \phi$ scalars which are the  gauge field components along $S^1$. We expand the fields $\vec \phi$ and $\vec \sigma$ as $\vec \phi=\frac{2\pi}{N_c}\rho+\frac{g^2}{4\pi}\vec b'$, and $\vec \sigma+\frac{\theta \vec \phi}{2\pi}=\frac{2\pi k+\theta}{N_c}+\sigma'$, where $g$ is the coupling constant, and the primed fields are the perturbations of  $\vec \phi$ and $\vec \sigma$ about their supersymmetric ground state. The ground state is labeled by an integer $k=0,...,N_c-1$, the vacuum angle $\theta$, and the Weyl vector $\vec \rho$. For more details regarding the group theoretic definitions and normalization, see \cite{Poppitz:2012nz}. The gauge holonomy (Wilson loop around $S^1$) in terms of the primed fields is $\Omega=\exp\left(i\frac{2\pi}{N_c}\vec H\cdot\vec\rho+i\frac{g^2}{4\pi} \vec H\cdot \vec b'\right)$, where $\vec H$ are the Cartan generators $\vec H=\left(H_1,...,H_{N_c-1}\right)$. At $\vec b'=0$ (supersymmetric vacuum), $\Omega$ obeys $\mbox{tr}\Omega=0$. We note that $\mbox{tr}\Omega$ is an order parameter of the $Z_{N_c}$ global center symmetry of the $SU(N_c)$ theory which is unbroken in the confining phase. Thus, we have $\mbox{tr}\Omega=0$ ($\mbox{tr}\Omega \ne 0$) in the confining (deconfienment) phase.  The total Lagrangian governing the system is given by Eq. (\ref{main lagrangian}), with potential (\ref{np potential}) which includes nonperturbative effects coming from monopole-instantons and bions. By studying this potential, we can determine the critical value of $L$ at which the transition happens.

Using the above mentioned conjecture, the authors in  \cite{Poppitz:2012nz} studied the phase transition of the pure $SU(N_c)$ theories for all groups $N_c>2$ in the absence of the vacuum angle $\theta$. In this letter, we retain the vacuum angle, and study its influence on the critical temperature $T^{\mbox{\scriptsize cr}}$  as well as  strength of  phase transition. We find that $T^{\mbox{\scriptsize cr}}$ is a periodic function of $\theta$ with period $2\pi$, and is analytic everywhere except at $(2n+1)\pi\,, n \in Z$, as expected for  physical observables \cite{Witten:1998uka}. For $\theta$ in the range $[0,\pi)$, the critical temperature decreases as $\theta$ increases. This finding is in agreement with recent lattice studies  that examined the dependence of the deconfinement temperature on $\theta$ \cite{D'Elia:2012vv}. On the other hand,  the strength of the transition, quantified as the change of  the trace of the Wilson line, increases with $\theta$. For large $N_c$, we find a weaker dependence of the deconfinement temperature and transition strength on the vacuum angle, due to  $N_c$ suppression.

%%%%%%%%%%%%%%%%%%%%%%%%%%%%%%%%%%%%%%%%%%%%%%%%%%%%%%%%%%%%%%%%%%%%%%%%%
%\section{The bion and monopole potentials and limit of metastability}
%%%%%%%%%%%%%%%%%%%%%%%%%%%%%%%%%%%%%%%%%%%%%%%%%%%%%%%%%%%%%%%%%%%%%%%%%
%\newpage
{\bf The bion and monopole potentials and limit of metastability:}  The system effective Lagrangian follows from the superpotential as given in \cite{Davies:2000nw}, and describes the dynamics of $SU(N_c)$ gauge group broken down into abelian $U(1)^{N_c-1}$ subgroups upon compactifying on $S^1$
\begin{eqnarray}
\nonumber
{\cal L}&=&\frac{1}{12\pi}\frac{m_W}{\log(m_W/\Lambda)}\left(\left(\partial_\mu \vec b'\right)^2+\left(\partial_\mu \vec \sigma'\right)^2\right)\\
&&+V_{\mbox{\scriptsize np}}+V_{\mbox{\scriptsize pert}}\,,
\label{main lagrangian}
\end{eqnarray}
where $\Lambda$ is the strong coupling scale, and $V_{\mbox{\scriptsize np}}$ and $V_{\mbox{\scriptsize pert}}$ are respectively the nonperturbative and perturbative potentials. The perturbative part comes from Gross-Pisarski-Yaffe one-loop effective action for the holonomy \cite{Gross:1980br}. As was shown in \cite{Poppitz:2012nz}, $V_{\mbox{\scriptsize pert}}$ is suppressed by three powers of $\log(m_W/\Lambda)$ compared to  $V_{\mbox{\scriptsize np}}$, and hence we neglect its contribution in our analysis. In the weak coupling limit, $\frac{g^2N_c}{4\pi}<<1$, the nonperturbative potential is due to bions and monopole-instantons \cite{Unsal:2007jx}, and is given by 
\begin{eqnarray}
\nonumber
\frac{V_{\mbox{\scriptsize np}}}{V^0_{\mbox{\scriptsize bion}}}&=&\sum_{i=1}^{N_c}e^{-2\vec \alpha_i\cdot \vec b'}-e^{-\left(\vec \alpha_i+\vec \alpha_{i+1}\right)\cdot \vec b'}\cos\left[\left(\alpha_i-\alpha_{i+1}\right)\cdot \vec\sigma' \right] \\
&-&c_m\sum_{i=1}^{N_c}e^{-\vec \alpha_i\cdot \vec b'}\cos\left[\vec \alpha_i\cdot \vec\sigma'+\psi \right]\,,
\label{np potential}
\end{eqnarray}
where $\psi=\frac{2\pi k+\theta}{N_c}$, and the parameter $k=0,1,..,N_c-1$ labels the vacuum branch, i.e. the branch with minimum ground energy.  We also introduced the dimensionless gaugino mass parameter $c_m=V^0_{\mbox{\scriptsize mon}}/V^0_{\mbox{\scriptsize bion}}$.  The  bion and monopole coefficients $V^0_{\mbox{\scriptsize bion}}$ and  $V^0_{\mbox{\scriptsize mon}}$, expressed in terms of the physical mass $m_W$ and the strong scale  $\Lambda$, are given by $V^0_{\mbox{\scriptsize bion}}=\frac{27}{8\pi} \frac{\Lambda^6}{m_W^3}\log \left(\frac{m_W}{\Lambda}\right)$, and $V^0_{\mbox{\scriptsize mon}}=\frac{9}{2\pi}\frac{m\Lambda^3}{m_W}\log\left(\frac{m_W}{\Lambda}\right)$.

First, we study the mass spectrum of the theory in the presence of the gaugino mass $m$ and the $\theta$ angle.  To this end, we expand $V_{\mbox{\scriptsize np}}$ to quadratic order in $\vec b'$ and $\vec \sigma'$. We also rescale $\vec b'$ and $\vec \sigma'$ as $\{b'^2,\sigma'^2\} \rightarrow \frac{6\pi\log\left(m_W/\Lambda\right)}{m_W} \{b'^2,\sigma'^2\} $ to have a canonically normalized Lagrangian. Then, the mass term can be diagonalized by using the $N_c$-dimensional root vectors $\vec \alpha_i= (0,0,...,\underbrace{1}_{i},\underbrace{-1}_{i+1},\underbrace{0}_{i+1},...)$ for $i=1,2,...N_c-1$ and $\vec \alpha_{N_c}=(-1,0,0,...,1)$. This results in an extra massless particle that decouples from the rest of the spectrum. Performing these steps we obtain
\begin{eqnarray}
\nonumber
V_{\mbox{\scriptsize np}}^{\mbox{\scriptsize quad}}&=&
 \frac{m_0^2}{2}\sum_{i=1}^{N_c}\left[\left(b'_{i+2}-2b'_{i+1}+b'_i\right)^2   \right.\\
\nonumber
&&\left.\quad+ \left(\sigma'_{i+2}-2\sigma'_{i+1}+\sigma'_i\right)^2 \right.\\
\nonumber
&&\left.\quad +c_m \left(\left(\sigma'_i-\sigma'_{i+1}\right)^2- \left(b'_i-b'_{i+1}\right)^2 \right)\cos\psi \right.\\
&&\left.\quad -2c_m\left(\sigma'_i-\sigma'_{i+1}\right)\left(b'_i-b'_{i+1}\right)\sin\psi\right]\,,
\end{eqnarray}
where $m_0^2=\frac{81}{4}\frac{\Lambda^6\left[\log\left(M_W/\Lambda\right) \right]^2}{m_W^4}$, and we notice that the monopole potential causes a mixing between the $\vec b'$ and $\vec \sigma'$ fields in the presence of a non-vanishing $\theta$. To  proceed further, we use the  $Z_{N_c}$ discrete  Fourier transform $\left\{\begin{array}{c} b'_j\\ \sigma'_j \end{array}  \right\}=\frac{1}{\sqrt{N_c}}\sum_{p=0}^{N_c-1} \left\{\begin{array}{c} \tilde b_p\\ \tilde\sigma_b \end{array}  \right\}e^{-2\pi i \frac{p_j}{N_c}}$ to find that the quadratic part of the bion potential is diagonal in the $\{\tilde b_p,\tilde \sigma_p\}$ basis, while the quadratic part of the monopole potential has both a diagonal component $\propto \cos\psi$ as well as an off-diagonal component $ \propto\sin\psi$ in the mass matrix. By performing an orthogonal transformation, we can go to new basis where the mass matrix is diagonal. The square of the mass spectrum in the new basis, as a function of $m$, is $m_{\pm}^2=16m_0^2\left[\sin^4\left(\frac{\pi p}{N_c}\right)\pm \frac{c_m}{4}\sin^2\left(\frac{\pi p}{N_c}\right)\right]$ for $p=1,...,N_c-1$, and we have dropped the unphysical massless mode $p=0$. We observe that the mass spectrum is independent of the $\theta$ angle. While this is true in the present case, restoring the $g^2 N_c$ term in the monopole potential in Eq. \ref{np potential}  will cause $m_{\pm}^2$ to have an explicit dependence on $\theta$, as was shown in the $SU(2)$ case in \cite{Poppitz:2012nz}. Considering the values of $m_{-}^2$, we see that the center-symmetric vacuum becomes locally unstable for $c_m>c_m^*$, where taking $p=1$, we find $c_m^*=4\sin^2\left(\frac{\pi}{N_c}\right)$ which is  independent of $\theta$. This corresponds to local instability at $L^*=\Lambda^{-1}\sqrt{\frac{4m}{3\Lambda}}\frac{1}{\frac{N_c}{\pi}\sin \frac{\pi}{N_c}} $ as one varies $L$ for a fixed value of the guagino mass $m$.

For any $N_c>2$ the non-perturbative potential contains cubic terms, and hence the transition is first order. In this case,  $c_m^*$ is the metastability limit above which the center-symmetric vacuum fails to be a local or global minimum. Unlike the metastable point $c_m^*$, the true critical value $c_m^{\mbox{\scriptsize cr}}$ depends on the value of the $\theta$ angle, and is always smaller than $c_m^*$.  
 
%%%%%%%%%%%%%%%%%%%%%%%%%%%%%%%%%%%%%%%%%%%%%%
%\section{Phase transition in $SU(3)$}
%%%%%%%%%%%%%%%%%%%%%%%%%%%%%%%%%%%%%%%%%%%%%%
%\newpage
{\bf Phase transition in $SU(3)$:} Before studying the phase transition, we must first determine the correct theory.  Since for a given $\theta$ we have $k=0,1,...,N_c-1$ branches, the correct theory is determined by selecting the branch with the minimum vacuum energy. The theta dependence of the vacuum energy density is ${\cal E}=\mbox{min}_{k}\left[-\frac{V^0_{\mbox{\scriptsize mon}}}{L}\cos\left(\frac{2\pi k+\theta}{N_c}\right) \right]$. For example, taking $N_c=3$  the vacuum branch is $k=0$ for $\theta \in (-\pi,\pi)$, and $k=2$ for $\theta \in (\pi, 3\pi)$. This implies that at $\theta=\pi$, the $k=0$ and $k=2$ branches are degenerate \cite{Witten:1998uka,Unsal:2012zj} (also, see  \cite{Thomas:2011ee} for interesting discussions about the $\theta$ term in a similar class of theories).

%%%%%%%%%%%%%%%%%%%%%%%%%
\begin{figure}[ht]
\centerline{
\includegraphics[width=.35\textwidth]{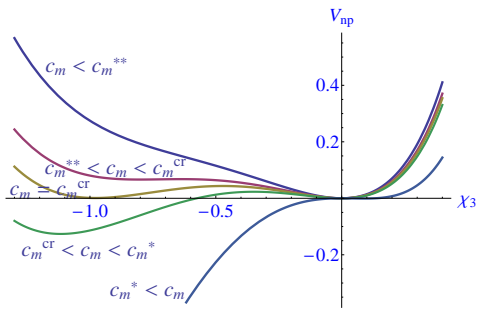}
}
\caption{The $N_c=3$ and $\theta=\pi/2$ effective potential $V_{\mbox{\scriptsize np}}$ (in units of $V_{\mbox{\scriptsize bion}}^0$)  as a function of $\chi_3$ for $\chi_2=\chi_4=0$, and $\chi_1\cong 0$.}
\label{su3VNP}
\end{figure}
%%%%%%%%%%%%%%%%%%%%%%%%%%%%%%

Now, we turn to the study of the phase transition in $SU(3)$. For  $\theta=0$, the $\vec b'$ and $\vec \sigma'$ fields do not mix to quadratic order. In this case, the expectation values of $\vec \sigma'$ and $b_2'$ remains zero, and one can plot the potential $V_{\mbox{\scriptsize np}}$ (a 2-dimensional plot) as a function of the deviation of the remaining holonomy from its center-symmetric value, i.e. $b_1'$, for different values of $c_m<c_m^*$.
Since $c^*_m$ is the metastable limit, we can plot the potential for a range of $c_m \leq c_m^*=4\sin^2\left(\frac{\pi}{3}\right)=3$ until we spot the true critical value $c_m^{\mbox{\scriptsize cr}}$ at which the first-order transition takes place. Turning on a non-zero value for $\theta$, will result in an admixture between $\vec \sigma'$ and $\vec b'$ fields at the quadratic level, which does not enable us to see the phase transition in 2-dimensional plots. To overcome this problem, we perform, as we mentioned above, an orthogonal transformation that brings us to new basis $\chi_{1,2,3,4}$ which are disentangled at the quadratic level. For $\theta \in \left(-\pi, \pi  \right)$, the vacuum branch is $k=0$, and the set of transformations is  $\left\{\begin{array}{c} b_1' \\b_2' \end{array} \right\}=- \left\{\begin{array}{c} \chi_1 \\ \chi_2 \end{array} \right\}\sin\left(\frac{\theta}{6}\right)- \left\{\begin{array}{c} \chi_3 \\ \chi_4 \end{array} \right\}\cos\left(\frac{\theta}{6}\right)$, and $\left\{\begin{array}{c} \sigma_1' \\\sigma_2' \end{array} \right\}= \left\{\begin{array}{c} \chi_1 \\ \chi_2 \end{array} \right\}\cos\left(\frac{\theta}{6}\right)- \left\{\begin{array}{c} \chi_3 \\ \chi_4 \end{array} \right\}\sin\left(\frac{\theta}{6}\right)$.
Now, the critical value $c_m^{\mbox{\scriptsize cr}}$, is obtained by plotting $V_{\mbox{\scriptsize np}}$ against $\psi_3$ for a range of $c_m<c_m^*=3$. This method is confirmed to a high precision  for  values of $|\theta| \lesssim \pi/2$, by comparing it to the results of an optimization algorithm (errors less than $0.5 \%$). Using this algorithm, we find that the expectation values of $\chi_2$ and $\chi_4$ remain zero, while $\chi_1$ and $\chi_3$ get expectation values such that $|\chi_3|/|\chi_1|>>1$. As $|\theta|$ increases beyond $\pi/2$ we find that $|\chi_1|$ gets expectation values comparable to those of $|\chi_3|$, and hence it becomes more difficult to see the first order transition on 2-dimensional plots. Instead of trying to apply a second orthogonal transformation that disentangles $\chi_1$ and $\chi_3$ in the case of large values of $|\theta|$, we use the same optmization algorithm, that is checked for small values of $|\theta|$, to determine the true minimum of the potential. Multiple checks were carried out to confirm our results.

The situation for $SU(3)$ and $\theta=\pi/2$ is shown in FIG. (\ref{su3VNP}), and quite identical to $\theta=0$ case.  Defining $c_m^{**}$ as the value of $c_m$ above which new metastable minima appears, we find that for $c_m<c_m^{**}$ there is a unique center-symmetric global minimum (confining phase).  For $c_m^{**}<c_m <c_m^{\mbox{\scriptsize cr}}$, there are three center-broken  metastable local minima (only one is shown, the other two are obtained by applying $Z_3$ rotations in the $\chi_3-\chi_4$ plane), and a unique center-symmetric global minimum (confining phase).  For $c_m^{\mbox{\scriptsize cr}}<c_m<c_m^*=3$, the center-symmetric  phase is a local minimum, and the three $Z_3$ breaking minima are global degenerate minima.  Finally, for $c_m>c_m^*$, the center-symmetric point fails to be a local minimum.
% In the right panel of FIG. (\ref{su3VNP}), we  plot the nonperturbative potential against $\chi_3$ at the corresponding critical values $c_m^{\mbox{\scriptsize cr}}$  for $\theta=0$ and $\theta=\pi/2$. We see that $|\chi_3^{\mbox{\scriptsize cr}}|$ increases as $\theta$ increases. 

%%%%%%%%%%%%%%%%%%%%%%%%%%%%%%%%%%%%%%%%%%%%%%%%%%%%%%%%%%%%%%%%%%%%%%%%%%%%%%%%%%%%%
%\section{The critical temperature and strength of transition in pure Yang-Mills} 
%%%%%%%%%%%%%%%%%%%%%%%%%%%%%%%%%%%%%%%%%%%%%%%%%%%%%%%%%%%%%%%%%%%%%%%%%%%%%%%%%%%%%%

%%%%%%%%%%%%%%%%%%%%%%%%%
\begin{figure}[ht]
\centerline{
\includegraphics[width=.3\textwidth]{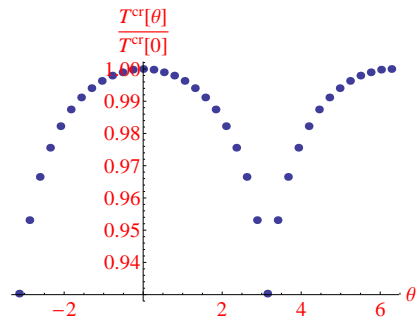}
}
\caption{The critical  normalized deconfinement temperature $\frac{T^{\mbox{\scriptsize cr}}(\theta)}{T^{\mbox{\scriptsize cr}}(0)}$.  The vacuum branch is $k=0$ for $\theta \in (-\pi,\pi)$, and $k=2$ for $\theta \in (\pi, 3\pi)$. We find that  $\frac{T^{\mbox{\scriptsize cr}}(\theta)}{T^{\mbox{\scriptsize cr}}(0)}$ is a periodic function of $\theta$ with period $2\pi$, and cusps at $\theta=(2n+1)\pi$ for any integer $n$, where two branches become degenerate. }
\label{su3 cr  theta}
\end{figure}
%%%%%%%%%%%%%%%%%%%%%%%%%%%%%%

%%%%%%%%%%%%%%%%%%%%%%%%%
\begin{figure}[ht]
\centerline{
\includegraphics[width=.3\textwidth]{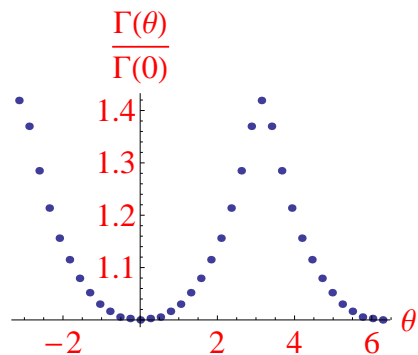}
}
\caption{The normalized discontinuity of the holonomy  $\Gamma(\theta)/\Gamma(0)=\Delta\left|\frac{g^2}{4\pi N_c}\mbox{tr}\Omega(\theta)\right|/\Delta\left|\frac{g^2}{4\pi N_c}\mbox{tr}\Omega(0)\right|$. We take $\theta$ in the range $ (-\pi, 3\pi)$, and $N_c=3$. The vacuum branch is $k=0$ for $\theta \in (-\pi,\pi)$, and $k=2$ for $\theta \in (\pi, 3\pi)$. We find that  $\Gamma(\theta)/\Gamma(0)$ is a periodic function of $\theta$ with period $2\pi$, and cusps at $\theta=(2n+1)\pi$ for any integer $n$, where two branches become degenerate. }
\label{su3 cr  theta 2}
\end{figure}
%%%%%%%%%%%%%%%%%%%%%%%%%%%%%%

{\bf The critical temperature and strength of transition in pure Yang-Mills:} Now, we are in a position to determine the critical temperature for pure $SU(3)$ Yang-Mills  as a function of $\theta$. To this end, we use the conjecture in \cite{Poppitz:2012sw,Poppitz:2012nz}  that the thermal phase transition in pure Yang-Mills theories  is continuously connected to a quantum phase transition in softly broken ${\cal N}=1$ super Yang-Mills on $R^{1,2} \times S^1$.  Using the relation $c_m=V^0_{\mbox{\scriptsize mon}}/V^0_{\mbox{\scriptsize bion}}=16\pi^2m/(3\Lambda^3L^2N_c^2)$, we see that at a fixed value of the gaugino mass $m$, the critical temperature $T^{\mbox{\scriptsize cr}}=L^{-1\mbox{\scriptsize cr}}$ is given by $T^{\mbox{\scriptsize cr}}(\theta)=T_0\sqrt{c_m^{\mbox{\scriptsize cr}}(\theta)}$, where $T_0=\sqrt{3\Lambda^3N_c^2/(16\pi^2m)}$. In FIG. \ref{su3 cr  theta}, we plot the normalized value of the critical temperature $T^{\mbox{\scriptsize cr}}(\theta)/T^{\mbox{\scriptsize cr}}(0)$ as a function of the vacuum angle $\theta$ in the range $(-\pi,2\pi)$. The critical temperature is a periodic function with period $2\pi$, and cusps at odd values of $\pi$ where two branches become degenerate and the $CP$ symmetry is spontaneously broken \cite{Witten:1998uka}. The normalized critical temperature decreases as $\theta$ increases in the range $(0,\pi)$. This is attributed to topological interference. The physics is that the effect of monopole-instantons is suppressed at $\theta=\pi$ with respect to $\theta=0$ by destructive interference among path histories. For small values of $\theta$ ($\theta \lesssim  1$), we fitted the ratio $\frac{T^{\mbox{\scriptsize cr}}(\theta)}{T^{\mbox{\scriptsize cr}}(0)}$ to the function  $1-{\cal R}^{\mbox{\scriptsize th}}\theta^{{\cal C}}$,  to find ${\cal R}^{\mbox{\scriptsize th}}=0.003$, and ${\cal C}=2.09$.  

 Such behavior is in excellent agreement with recent lattice studies that examined the $\theta$ dependence of the deconfinement temperature in $SU(3)$ pure Yang-Mills \cite{D'Elia:2012vv} (also, see \cite{Bonati:2013tt} for a study of the $\theta$ dependence across the deconfining transition, and also  \cite{Lucini:2012gg} for a review ). In the presence of $\theta$, the Euclidean path integral representation of the partition function is not suitable for Monte-Carlo simulations because the measure is complex. The authors in \cite{D'Elia:2012vv}  overcame this problem by studying the theory at imaginary $\theta$, where the problem disappears, and then made use of analytic continuation to infer the dependence at real $\theta$. By doing that, they were able to extract the $\theta$ dependence of $T^{\mbox{\scriptsize cr}}$ for small values of $\theta$. They found $\left[\frac{T^{\mbox{\scriptsize cr}}(\theta)}{T^{\mbox{\scriptsize cr}}(0)}\right]^{\mbox{\scriptsize lattice}}=1-{\cal R}^{\mbox{\scriptsize lattice}}\theta^2$, with ${\cal R}^{\mbox{\scriptsize lattice}}=0.0175$. This is  $5$ times larger than the value of ${\cal R}^{\mbox{\scriptsize th}}$. However, this should not come as a surprise since in general one does not expect an exact numerical match between a weakly coupled setup and pure Yang-Mills, a strongly coupled theory. 

The discontinuity at the first order phase transition can be quantified in terms of the change in the trace of the Wilson line. In the weak limit, $g^2N_c/4\pi<<1$, we have $\mbox{tr}\Omega(\theta)\cong i\left(\frac{g^2N_c}{4\pi}\right)\frac{1}{N_c}\mbox{tr}\left(\Omega_0\vec H\cdot \vec b'(\theta)\right)$, where $\Omega_0=\exp\left[{i\frac{2\pi}{N_c}\vec H\cdot \vec\rho}\right]$ is the center-symmetric holonomy. In FIG. \ref{su3 cr  theta 2}, we plot the discontinuity of the absolute value of the holonomy, normalized to unity at $\theta=0$, as a function of $\theta$ for $N_c=3$. We find that the change in the trace of the Wilson line has an opposite behavior to that of the critical temperature; it increases with increasing $\theta$. For small values of $\theta$, we can fit $\Delta\left|\mbox{tr}\Omega(\theta)\right|/\Delta\left|\mbox{tr}\Omega(0)\right|$ to the function $1+{\cal R}\theta^{\cal C}$ with ${\cal R}=0.03$, and ${\cal C}=2$. This behavior indicates that the strength of the phase transition is enhanced for non-zero values of the vacuum angle. It is interesting to see if this trend can also be observed in lattice simulations by generalizing the study in \cite{Mykkanen:2012ri} to non zero values of $\theta$. 

  We can also define the "latent heat" $\Delta Q$ of the system by identifying the quantity $V_{\mbox{\scriptsize np}}/L$ as the free energy density  $F$, and then employing the thermodynamic relation $F=E-TS$, where $E$ is the internal energy density,  $T=1/L$, and $S=-\partial F/\partial T$ is the entropy density. Now, using $T=T_0\sqrt{c(\theta)}$, and realizing that the latent heat is the difference between the internal energy at the symmetric and broken phases, we can show that $\frac{\Delta Q(\theta)}{\Delta Q(0)}=\left(\frac{c_m^{\mbox{\scriptsize cr}}(\theta)}{c_m^{\mbox{\scriptsize cr}}(\theta=0)}\right)^{3/2}\frac{\left|V_m\left(\vec \sigma'^{\mbox{\scriptsize cr}},\vec b'^{\mbox{\scriptsize cr}},\theta\right)-V_m(\vec 0,\theta)\right|}{\left|V_m\left(\vec \sigma'^{\mbox{\scriptsize cr}},\vec b'^{\mbox{\scriptsize cr}},\theta=0\right)-V_m(\vec 0,\theta=0)\right|}$, where $V_m\left(\vec \sigma',\vec b',\theta\right)=\sum_{i=1}^{N_c}e^{-i\vec \alpha_i \cdot \vec b'}\cos\left[\vec\alpha_i\cdot \vec \sigma'+\psi\right]$, and $\vec \sigma'^{\mbox{\scriptsize cr}}$, $\vec b'^{\mbox{\scriptsize cr}}$ are the values of these quantities in the broken phase. We  studied the quantity $\Delta Q(\theta)/\Delta Q(0)$, to find that it has the exact same behavior of the trace of the Wilson line.       

\begin{table}
\centerline{
\begin{tabular}{|c||c|c||c|c||c|c||}
\hline
    & $N_c=6$&  $N_c=6$&   $N_c=7$&  $N_c=7$& $N_c=8$&  $N_c=8$  \\ \hline
   $\theta$  &   $c_m^{\mbox{\scriptsize cr}}$ & $\Gamma$& $c_m^{\mbox{\scriptsize cr}}$ & $\Gamma$ &$c_m^{\mbox{\scriptsize cr}}$ & $\Gamma$ \\ \hline\hline
 $0$    & 0.6739  & 0.9105  &   0.4956  & 1.0681 & 0.3792 & 1.2441 \\\hline
 $\pi/12$  & 0.6735  & 0.9107  &   0.4953  & 1.0680 & 0.3790 & 1.2447\\\hline
 $\pi/6$   & 0.6724  & 0.9114  &   0.4946  & 1.0683 & 0.3786 &1.2462 \\\hline
 $\pi/4$   & 0.6706 & 0.9128  &   0.4935  & 1.0689 & 0.3778 &1.2486\\\hline
 $\pi/3$   & 0.6680  & 0.9144  &   0.4918  & 1.0692 &0.3767 & 1.2517\\\hline
 $5\pi/12$  & 0.6646  & 0.9164  &   0.4896  & 1.0695 &0.3752 &1.2555 \\\hline
 $\pi/2$  & 0.6604  & 0.9186  &   0.4869  & 1.0696 &0.3735 &1.2596\\\hline 
 \end{tabular}
}
\caption{The critical values of the gaugino mass parameter $c_m^{\mbox{\scriptsize cr}}$, and the Wilson line discontinuity $\Gamma(\theta)=\Delta\left|\frac{g^2}{4\pi N_c}\mbox{tr}\Omega(\theta)\right|$  for a range of $\theta \in [0,\pi/2]$, and $N_c=6,7$, and $8$.  }
\label{critical values}
\end{table}

%%%%%%%%%%%%%%%%%%%%%%%%%%%%%%%%%%%%%%%%%%%%%%%%%%%%%%%%%%%%
%\section{The large $N_c$ limit and concluding remarks}
%%%%%%%%%%%%%%%%%%%%%%%%%%%%%%%%%%%%%%%%%%%%%%%%%%%%%%%%%%%%%

{\bf The large $N_c$ limit and concluding remarks:} For $N_c>3$, the nonperturbative potential $V^{\mbox{\scriptsize np}}$ is a function of $N_c-1$ holonomies. In this case, we use an optimization algorithm to find the minima of $V^{\mbox{\scriptsize np}}$.
 In table (\ref{critical values}) we list the critical values of the gaugino mass parameter $c_m^{\mbox{\scriptsize cr}}$ for a range of $\theta \in [0,\pi/2]$, and for  $N_c=6,7$, and $8$. Similar to $N_c=3$ case, we observe the same pattern of the $\theta$ dependence of the deconfinement temperature and  the Wilson line discontinuity. However, as $N_c$ increases, the dependence of $T^{\mbox{\scriptsize cr}}$ and $\mbox{tr}\Omega$ on $\theta$ becomes weaker. In the large $N_c$ limit, one expects that both quantities become independent of $\theta$. Using the values of $c_m^{\mbox{\scriptsize cr}}$ listed in table \ref{critical values}, we can fit  $\frac{T^{\mbox{\scriptsize cr}}(\theta)}{T^{\mbox{\scriptsize cr}}(0)}$ to the function $1-{\cal R}_2\theta^2/N^{{\cal D}_2}-{\cal R}_4\theta^4/N^{{\cal D}_4}$ to find ${\cal R}_2=0.021$, ${\cal R}_4=0.0036$, ${\cal D}_2=0.93$, and ${\cal D}_4=2.36$. However, we noticed that the values of ${\cal D}_2$, and ${\cal D}_4$ get bigger as we restrict our fits to larger values of $N_c$. It was numerically challenging to extend our analysis beyond $N_c=8$ in order to check the stability of ${\cal D}_{2,4}$ as $N_c$ increases. 
 
In this letter, we have demonstrated the use of a class of Yang-Mills theories on $R^{1,2}\times S^1$ as an analytic tool to learn about realistic gauge field theories. Interestingly enough, this class of theories is under complete control, yet resembles the original ones.  It is the hope that, by contrasting  the analytic predictions with the lattice data, these models will open the path into a new paradigm to tackle strong coupling problems. This work, along with \cite{Poppitz:2012sw, Anber:2011gn,Poppitz:2012nz}, is a step toward exploring such a new path.

I would like to thank Erich Poppitz for enlightening discussions, and Mithat Unsal and Massimo D'Elia for interesting communications and  valuable comments on the manuscript. This work has been supported by NSERC Discovery Grant of Canada.

\end{document}